\documentclass[a4paper,12pt, epsfig]{article}
\def\letter{0}\def\pr{0}
\pdfoutput=1 %per jhep
\usepackage{epsfig}
\usepackage{epstopdf}
\usepackage{graphicx}
\usepackage{ifthen}
\usepackage{ulem}

\usepackage{amsmath}
\usepackage[psamsfonts]{amssymb}
\usepackage{euscript}

\usepackage{latexsym}
\usepackage[arrow,matrix,curve]{xy}

\usepackage[hypertexnames=false]{hyperref}
\hypersetup{
    colorlinks=true,
    linkcolor=blue,
    filecolor=magenta,   
    citecolor=black,   
    urlcolor=cyan,
    }

\newskip\humongous \humongous=0pt plus 1000pt minus 1000pt

\newif\ifdtup

\def\,{\hspace{-.1cm}}
\def\hsp{,\hspace{.7cm}}

\def\fc#1#2 {\frac{n}{q}#1\frac{n}{q}#2}

\def\kt{\kappa}
\def\kt{\mathfrak{K}}
\def\ks{|\kt\rangle}

\def\hp{H^{\prime(t)}}
\def\rv{{\rm{vac}}}

\newcommand{\vac}{\ensuremath{|0\rangle}}

\renewcommand{\sin}{\textrm{sin}}

\renewcommand{\sinh}{\textrm{sinh}}
\renewcommand{\cosh}{\textrm{cosh}}
\renewcommand{\tanh}{\textrm{tanh}}
\newcommand{\sech}{\textrm{sech}}
\newcommand{\csch}{\textrm{csch}}

\def\exp#1{\hbox{\rm exp}\left[#1\right]}

\def\sign#1{{\rm sign}\left(#1\right)}

\renewcommand{\theequation}{\arabic{section}.\arabic{equation}}
\renewcommand{\(}{\begin{equation}}
\renewcommand{\)}{end{equation} \vspace{-.05in}\linebreak}

\newcounter{saveeqn}
\newcounter{savealpheqn}

\newcommand{\alpheqn}{\setcounter{saveeqn}{\value{equation}}%
  \stepcounter{saveeqn}\setcounter{equation}{0}%
  \renewcommand{\theequation}{\mbox{\arabic{section}.\arabic{saveeqn}
\alph{equation}}}
  \renewcommand{\)}{\end{equation}}}
\def\part#1{\frac{\partial}{\partial{#1}}}%
\def\group#1{\refstepcounter{equation}\setcounter{saveeqn}
 {\value{equation}}%
  \label{#1}\setcounter{equation}{0}%
\renewcommand{\theequation}{\mbox{\arabic{section}.\arabic{saveeqn}
\alph{equation}}}
  \renewcommand{\)}{\end{equation}}}
\newcommand{\reseteqn}{\setcounter{equation}{\value{saveeqn}}%
  \renewcommand{\theequation}{\arabic{section}.\arabic{equation}}%
  \renewcommand{\)}{\end{equation}}}

\newcommand{\aalpheqn}{\setcounter{saveeqn}{\value{equation}}%
  \stepcounter{saveeqn}\setcounter{equation}{0}%
  \renewcommand{\theequation}{\mbox{
        \Alph{subsection}.\arabic{saveeqn}\alph{equation}}}
   \renewcommand{\)}{\end{equation}}}
\newcommand{\areseteqn}{\setcounter{equation}{\value{saveeqn}}%
  \renewcommand{\theequation}{\Alph{subsection}.\arabic{equation}}%
  \renewcommand{\)}{\end{equation}}}

\renewcommand{\thefootnote}{\alph{footnote}}
\renewcommand{\(}{\begin{equation}}
\renewcommand{\)}{\end{equation}}
\newcommand{\ba}{\begin{eqnarray}}
\newcommand{\ea}{\end{eqnarray}}

\renewcommand{\b}{\beta}

\renewcommand{\sl}{{\sqrt{\lambda}}}

\newcommand{\cbp}{\mathop{\vtop{\ialign{##\crcr
   $\hfil\displaystyle{}\hfil$\crcr\noalign{\kern-13pt\nointerlineskip}
   \BIG{)}\hskip0pt\crcr\noalign{\kern3pt}}}}}
\newcommand{\pa}{\mathop{\vtop{\ialign{##\crcr

$\hfil\displaystyle{\oplus}\hfil$\crcr\noalign{\kern+1pt\nointerlineskip
}
   \hspace{.08in}$^{\alpha=0}$\hskip6pt\crcr\noalign{\kern3pt}}}}}
\renewcommand{\hsp}{,\hspace{.3in}}
\newcommand{\p}{^\prime}
\newcommand{\pp}{^{\prime\prime}}

\catcode`\@=11
\def\vereq#1#2{\lower3pt\vbox{\baselineskip1.5pt \lineskip1.5pt
\ialign{$\m@th#1\hfill##\hfil$\crcr#2\crcr\sim\crcr}}}
\catcode`\@=12

\renewcommand{\(}{\begin{equation}}
\renewcommand{\)}{\end{equation}}

\def\pin#1{\int \frac{d#1}{2\pi}}
\def\ppin#1{\int\hspace{-17pt}\sum \frac{d#1}{2\pi}}
\def\ppink#1{\int\hspace{-17pt}\sum\frac{d^{#1}k}{(2\pi)^{#1}}}
\def\ppinkp#1{\int\hspace{-17pt}\sum\frac{d^{#1}k\p}{(2\pi)^{#1}}}
\def\dint{\int\hspace{-12pt}\sum }
\def\pink#1{\int \frac{d^{#1}k}{(2\pi)^{#1}}}

\def\pinkp#1{\int \frac{d^{#1}k\p}{(2\pi)^{#1}}}

\def\Bd#1{B^\ddag_{k_{#1}}}
\def\Btd#1{B^{(t)\ddag}_{k_{#1}}}
\def\Bt#1{B^{(t)}_{k_{#1}}}
\def\Bdp#1{B^\ddag_{k\p_{#1}}}

\def\cc{\mathcal{C}}
\def\df{\mathcal{D}_{f}}
\def\dft{\mathcal{D}_{f}^{(t)}}
\def\dfd{\mathcal{D}_{f}^{(t)\dag}}

\def\I{\mathcal{I}}

\def\os{\omega_S}

\def\gx{(\gamma(x-vt))}

\def\blu#1{\textcolor{blue}{Alberto: #1}}
\def\red#1{\textcolor{red}{Jarah: #1}}

\newcommand{\beas}{\begin{eqnarray*}}
\newcommand{\eeas}{\end{eqnarray*}}

\newcommand{\bquo}{\begin{quote}}
\newcommand{\enqu}{\end{quote}}

\def\lim#1{\stackrel{\rm{lim}}{{}_{#1}}}

\newcommand{\R}{{\mathbb R}}

\newcommand{\g}{\mathfrak g}

\def\ch{{\mathcal{H}}}
\def\co{{\mathcal{O}}}

\def\ok#1{\omega_{k_{#1}}}
\def\okp#1{\omega_{k\p_{#1}}}
\def\okt#1{\omega_{\kt_{#1}}}
\def\V#1{V^{(#1)}(\sqrt{\lambda}f(x))}
\def\Vg#1{V^{(#1)}(\sqrt{\lambda}f(\gamma(x-vt))}
\def\v#1{V^{(#1)}[f(x),x]}
\def\ck{\csch\left(\frac{\pi k}{2\b}\right)}

\def\mb{\mathcal{B}}
\def\mc{\mathcal{C}}
\def\md{\mathcal{D}}
\def\me{\mathcal{E}}

\newcommand{\beq}{\begin{equation}}
\newcommand{\eeq}{\end{equation}}
\newcommand{\bea}{\begin{eqnarray}}
\newcommand{\eea}{\end{eqnarray}}

\newskip\humongous \humongous=0pt plus 1000pt minus 1000pt

\newif\ifdtup

\catcode`\@=11

\ifthenelse{\equal{\letter}{0}}{ %se lettere=0, commenta questo

\@addtoreset{equation}{section}
\def\theequation{\arabic{section}.\arabic{equation}}

\def\@normalsize{\@setsize\normalsize{15pt}\xiipt\@xiipt
\abovedisplayskip 14pt plus3pt minus3pt%
\belowdisplayskip \abovedisplayskip
\abovedisplayshortskip \z@ plus3pt%
\belowdisplayshortskip 7pt plus3.5pt minus0pt}

\def\small{\@setsize\small{13.6pt}\xipt\@xipt
\abovedisplayskip 13pt plus3pt minus3pt%
\belowdisplayskip \abovedisplayskip
\abovedisplayshortskip \z@ plus3pt%
\belowdisplayshortskip 7pt plus3.5pt minus0pt
\def\@listi{\parsep 4.5pt plus 2pt minus 1pt
      \itemsep \parsep
      \topsep 9pt plus 3pt minus 3pt}}

\relax

\def\section{\@startsection{section}{1}{\z@}{3.5ex plus 1ex minus  .2ex}{2.3ex plus .2ex}{\large\bf}}

\def\thesection{\arabic{section}}
\def\thesubsection{\arabic{section}.\arabic{subsection}}

\def\appendix{\setcounter{section}{0}
 \def\thesection{Appendix \Alph{section}}
 \def\thesubsection{\Alph{section}.\arabic{subsection}}
 \def\theequation{\Alph{section}.\arabic{equation}}}
\renewcommand{\theequation}{\arabic{section}.\arabic{equation}}

}{
\renewcommand{\theequation}{\arabic{equation}}

} %se letter=0, commenta questo

\begin{document}
% ========================================================================
\def\thefootnote{\fnsymbol{footnote}}
\def\thetitle{Instanton Corrections to the MSTB Kink Mass}
\def\autone{Jarah Evslin}
\def\autthree{Alberto Garc\'ia Mart\'in-Caro}
\def\autfour{Andrzej Wereszczy\'nski}
\def\affc{Institute  of  Theoretical Physics,  Jagiellonian  University,  Lojasiewicza  11,  Krak\'ow,  Poland}
\def\affcc{International Institute for Sustainability with Knotted Chiral Meta Matter (WPI-SKCM2), Higashi-Hiroshima, Hiroshima 739-8526, Japan}
\def\affb{University of the Chinese Academy of Sciences, YuQuanLu 19A, Beijing 100049, China}
\def\affa{Institute of Modern Physics, NanChangLu 509, Lanzhou 730000, China}
\def\bila{Department of Physics, University of the Basque Country UPV/EHU, Bilbao, Spain}
\def\bilb{EHU Quantum Center, UPV/EHU, Bilbao, Spain}

%\title{Titolo}

\ifthenelse{\equal{\pr}{1}}{
\title{\thetitle}
\author{\autone}
\author{\auttwo}
\author{\autthree}
\author{\autfour}
\affiliation {\affa}
\affiliation {\affb}
%\affiliation {\affc}
%\affiliation {\affd}
%pr e uno

}{

\begin{center}
{\large {\bf \thetitle}}

\bigskip

\bigskip

%\catcode`@=11

{\large \noindent  \autone,${}^{1,2}$
\footnote{jarah@impcas.ac.cn} %{
%\auttwo{${}^{3}$}
%\footnote{alonsoiz@usal.es}
 \autthree{${}^{3,4}$}\footnote{alberto.garciam@ehu.eus}
\ and \autfour{${}^{5,6}$} \footnote{andwereszczynski@gmail.com}
}

%{\large \noindent  \autone{${}^{1,2}$} \footnote{jarah@impcas.ac.cn} and \auttwo{${}^{1,2}$} \footnote{guohengyuan@impcas.ac.cn}}

\vskip.7cm

1) \affa\\
2) \affb\\
%3) \sal\\
3) \bila\\
4) \bilb\\
5) \affc\\
6) \affcc\\

\end{center}

}

\begin{abstract}
\noindent
The Montonen–Sarker–Trullinger–Bishop (MSTB) model enjoys two classically degenerate kink solutions in the same topological sector.  We construct the instanton that interpolates between them and argue that the two lowest lying Hamiltonian eigenstates in the kink sector of the corresponding quantum theory correspond to symmetric and antisymmetric combinations of coherent states localized at these two solutions.  We use the instanton gas approximation to provide a simple, analytic formula for the leading contribution to the mass splitting between these two states.

\end{abstract}

% \vfill
%
% \end{titlepage}
\setcounter{footnote}{0}
\renewcommand{\thefootnote}{\arabic{footnote}}

\ifthenelse{\equal{\pr}{1}}
{
\maketitle
}{}

\section{Introduction}

Topological solitons in quantum field theories are generally approximated to be squeezed, coherent states plus perturbative corrections.  The states need to be coherent \cite{vinc72,cornwall74} so that the expectation value of the field is close to the classical solution, while the role of the squeezing \cite{cahill76,mekink,cocorr23} is so that discrete and continuum normal modes, rather than plane waves, are turned off in the ground state.  In the case of the domain wall soliton in 3+1 dimensions, the squeeze is necessary to remove an ultraviolet divergence in the wall's one-loop tension \cite{noi4d}, while in lower dimensions it leads to a finite decrease in the soliton's energy density.  Such an approach allows computations of masses, form factors, lifetimes and scattering amplitudes as a series expansion in the coupling.

However, even at weak coupling, this treatment is an approximation.  This is a result of excursions of the quantum field localized at one minimum of the potential to the basin of another minimum.  Such excursions are responsible for instanton corrections.  In the case of BPS solitons in supersymmetric theories, such instanton corrections are often the only corrections that are allowed by nonrenormalization theorems and they play a central role in the dynamics.  For example, in $\mathcal{N}=2$ supersymmetric QCD, instanton corrections to the monopole mass shift the location in moduli space where the monopole condenses, leading to expectation values that govern the symmetry breaking patterns of the vacua and even lead to confinement when some supersymmetry is softly broken \cite{sw1,sw2}.  This motivates an understanding of such instanton corrections beyond supersymmetric theories.

In the present paper, we introduce perhaps the simplest example of such an instanton correction.  As reviewed in Sec.~\ref{modsez}, the MSTB model \cite{mstb1,mstb2,mstb3} contains two kink solutions in the same topological sector, whose splitting is governed by a parameter $\alpha$.  Although the kink solutions are known analytically, as a warmup in Sec.~\ref{kinksez} we construct the kink solutions as an expansion in $\alpha$.  Next, in Sec.~\ref{instsez}, we construct the instanton that interpolates between them, again as an expansion in $\alpha$.  The calculation of instanton corrections thus becomes a higher-dimensional analogue of the double-well problem in quantum mechanics.  This allows us, in Sec.~\ref{azsez}, to compute the corresponding mass splitting using a standard instanton gas approach.  Our main result is that the leading contribution to the mass splitting is, up to prefactors which are polynomial in $\alpha$,
\beq
\Delta M\sim{\rm{exp}}\left(
-\frac{4m^2\alpha^3}{3\sqrt{2}\lambda}
\right)
\eeq
where $m$ is the mass of the fundamental meson and $\lambda\phi^4/4$ is the  quartic coupling.

\section{The MSTB Model} \label{modsez}

\subsection{The Hamiltonian}

The (1+1)-dimensional MSTB model contains two real Schrodinger picture scalar fields $\phi_1(x)$ and $\phi_2(x)$ together with their conjugate momenta $\pi_1(x)$ and $\pi_2(x)$.  It is described by the Hamiltonian
\beq
H=\int dx :\ch(x):\hsp
\ch(x)=\frac{\sum_{a=1}^2\left(\pi^2_a(x)+\partial_x \phi_a(x)\partial_x \phi_a(x)\right)}{2}+U(\phi_1(x),\phi_2(x))
\eeq
where $::$ is the usual normal ordering of plane waves and the potential is
\beq
U(\phi_1(x),\phi_2(x))=\frac{\lambda\left(\phi_1(x)^2+\phi_2(x)^2-\frac{m^2}{2\lambda}\right)^2}{4}+\frac{(1-\alpha^2)m^2}{8}\phi_2(x)^2
\eeq
which reduces to the form in Ref.~\cite{albertomstb} via the identifications $\lambda=m=2$, $\alpha^2=1-\sigma^2$.  

We will be interested in the double scaling limit for the dimensionless parameters
\beq
\frac{\lambda}{m^2}\rightarrow 0\hsp
\alpha\rightarrow 0^+\hsp
\frac{\alpha}{(\lambda/m^2)^\beta}\rightarrow\infty
\eeq
for all $\beta>0$.  Intuitively, $\alpha$ is very small but remains classical, while the expansion in the much smaller $\lambda/m^2=\lambda\hbar/m^2$ is equivalent to the semiclassical expansion.  In particular, the third condition allows for a reasonable semiclassical expansion about the vacua
\beq
\phi_1(x)=\pm v\hsp \phi_2(x)=0\hsp v=-\frac{m}{\sqrt{2\lambda}}.
\eeq
At each fixed order in the semiclassical expansion, we will consider the $\alpha$ expansion.  

While the small $\lambda$ limit is necessary for our approach, the small $\alpha$ limit is just a convenience so that we may obtain analytical results.  Later will also present numerical results at finite $\alpha$.

The corresponding classical field theory describes the fields $\phi_1(x,t)$ and $\phi_2(x,t)$ which obey the classical equations of motion
\bea
-\ddot \phi_1+\phi^{\pp}_1&=&\lambda\phi_1\left(\phi_1^2+\phi_2^2-\frac{m^2}{2\lambda}\right)\\
-\ddot \phi_2+\phi^{\pp}_2&=&\lambda\phi_2\left(\phi_1^2+\phi_2^2-\frac{m^2}{2\lambda}\right)+\frac{\left(1-\alpha^2\right)m^2}{4}\phi_2.\nonumber
\eea
These equations admit two time-independent, stable kink solutions in the same topological sector
\beq
\phi_i(x,t)=f^\pm_i(x)\hsp
f_1^\pm=\frac{m}{\sqrt{2\lambda}}\tanh{}\left(\sqrt{1-\alpha^2}\frac{mx}{2}\right)\hsp f_2^\pm=\pm\alpha\frac{m}{\sqrt{2\lambda}}\ \sech\left(\sqrt{1-\alpha^2}\frac{mx}{2} \right) \label{es}
\eeq
where for concreteness we have fixed the center of mass at $x=0$.  Note that at $\alpha=0$, which does not satisfy our double scaling limit, these two solutions merge into the usual kink of the $\phi^4$ double-well model.

\subsection{The Expansion}

The instanton is a solution $\phi_i(x,t)=F_i(x,t)$ of the Euclidean time equations of motion
\bea
\ddot F_1+F_1^{\pp}&=&\lambda F_1\left(F_1^2+F_2^2-\frac{m^2}{2\lambda}\right)\label{eom}\\
\ddot F_2+F^{\pp}_2&=&\lambda F_2\left(F_1^2+F_2^2-\frac{m^2}{2\lambda}\right)+\frac{\left(1-\alpha^2\right)m^2}{4}F_2\nonumber
\eea
which interpolates between the two stable kinks
\beq
\lim{t\rightarrow \pm\infty}F_i(x,t)=f_i^\pm(x) .\label{tbc}%\hsp
%\lim{t\rightarrow \infty}F_i(x,t)=f_i^+(x).
\eeq
We also impose the boundary conditions
\beq
F_1(\pm\infty,t)=\pm \frac{m}{\sqrt{2\lambda}}\hsp 
F_2(\pm\infty,t)=0. \label{xbc}
\eeq

As $\alpha<<1$, we will expand the instanton solution $F_i$ as a power series in $\alpha$.  In particular, we will expand $F_1(x,t)$ about the usual $\phi^4$ kink solution
\beq
F_1(x,t)=F_1^{(0)}(x,t)+\alpha^2 F_1^{(2)}(x,t)+\alpha^4 F_1^{(4)}(x,t)+O(\alpha^6)\hsp 
F_1^{(0)}(x,t)=\frac{m}{\sqrt{2\lambda}}\tanh{}\left(\frac{mx}{2} \right) \label{ordz}
\eeq
and we will expand $F_2(x,t)$ about zero
\beq
F_2(x,t)=\alpha F_2^{(1)}(x,t)+\alpha^3 F_2^{(3)}(x,t)+O(\alpha^5)\hsp
\lim{t\rightarrow \pm\infty}F_2^{(1)}(x,t)=\pm\frac{m}{\sqrt{2\lambda}}\ \sech\left(\frac{mx}{2} \right).
\eeq

\subsection{Order $O(\alpha)$}

The second Euclidean equation of motion (\ref{eom}) then becomes, at order $O(\alpha)$
\beq
\ddot F^{(1)}_2\big|_{O(\alpha^0)}+F^{(1)\prime\prime}_2=\lambda F^{(1)}_2\left(F^{(0)2}_1-\frac{m^2}{4\lambda}\right)=\frac{m^2}{4}\left(1-2\sech^2\left(\frac{mx}{2}\right)
\right)F^{(1)}_2.
\eeq
Note that this is a homogeneous equation for $F^{(1)}_2$ and so does not yet fix its normalization, although at large $|t|$ it is fixed by the boundary conditions.  At general $t$, the normalization can only be fixed at order $O(\alpha^3)$, as is true even in the time-independent case.

Let us assume that all time evolution is slow, so that each time derivative introduces at least one power of $\alpha$.  This is plausible for the instanton, as it starts stationary and it only moves a distance $\alpha$.  As a result, the $\ddot F^{(1)}_2$ term does not contribute at order $O(\alpha)$.  

Imposing the boundary conditions (\ref{xbc}), this homogeneous ordinary differential equation is easily solved
\beq
F^{(1)}_2(x,t)=A(t)\sech\left(\frac{mx}{2}\right)
\eeq
for some function $A(t)$ which is constrained by Eq.~(\ref{tbc}) to obey
\beq
\lim{t\rightarrow \pm\infty}A(t)=\pm \frac{m}{\sqrt{2\lambda}}.
\eeq

\section{Kinks}  \label{kinksez}

To better understand the expansion, let us momentarily forget about Eq.~(\ref{tbc}) and study time-independent solutions
\beq
F_i(x,t)=f_i(x)\hsp
f_1^{(0)}(x)=\frac{m}{\sqrt{2\lambda}}\tanh{}\left(\frac{mx}{2} \right).
\eeq
Thus, in this subsection we will write $A(t)=A$ where $A$ is an arbitrary constant.   %Of course we already know the exact solution (\ref{es}) for all values of $\alpha$, but as a warm up for the instanton calculation it will be useful to derive the first few orders.

\subsection{Order $O(\alpha^2)$}

At order $O(\alpha^2)$, the first equation in Eq.~(\ref{eom}) becomes
\bea
%-\ddot f_1^{(2)}
f_1^{(2) \pp}&=&\lambda f_1^{(2)}\left(3f_1^{(0)2}-\frac{m^2}{2\lambda}\right)+\lambda F^{(0)}_1f_2^{(1)2}\\
&=&m^2 f_1^{(2)}\left(1-\frac{3}{2}\sech^2\left(\frac{mx}{2}\right)\right)+m\sqrt{\frac{\lambda}{2}} A^2 \sech^2 \left(\frac{mx}{2}\right)\tanh \left(\frac{mx}{2}\right)\nonumber
\eea
which is solved by
\beq
f_1^{(2)}=-\sqrt{\frac{\lambda}{2}}\frac{A^2}{2} x\ \sech^2\left(\frac{mx}{2}\right)+c_1^{(2)}\sech^2\left(\frac{mx}{2}\right).
\eeq
Here the constant of integration $c_1^{(2)}$ corresponds to the kink's zero mode.  We can shift the origin of $x$ to set $c_1^{(2)}=0$. The first term is proportional to the Derrick mode of the kink in the decoupled limit $f_1^{(0)}$. This is not surprising as a non-zero value of $\alpha$ compresses the kink. 

\subsection{Order $O(\alpha^3)$}

Next, we proceed to the second equation at order $O(\alpha^3)$
\bea
f^{(3) \pp}_2&=&\lambda f^{(3)}_2\left(f_1^{(0)2}-\frac{m^2}{2\lambda}\right)
+\lambda f_2^{(1)}\left(2f_1^{(0)}f_1^{(2)}+f_2^{(1)2}\right)
+\frac{m^2}{4}\left(f_2^{(3)}-f_2^{(1)}\right)\nonumber\\
%&=&\frac{m^2}{4}\left(1-2\sech^2 \left(\frac{mx}{2}\right)
%\right)F^{(3)}_2-\frac{m^2 A}{4}\sech\left(\frac{mx}{2}\right)\nonumber\\
%&&+\lambda A\sech\left(\frac{mx}{2}\right)
%\left( \frac{A^2}{2}x\tanh\left(\frac{mx}{2}\right)\sech^2\left(\frac{mx}{2}\right)
%+A^2\sech^2\left(\frac{mx}{2}\right)\right)\nonumber\\
&=&\frac{m^2}{4}\left(1-2\sech^2 \left(\frac{mx}{2}\right)
\right)F^{(3)}_2-\frac{m^2 A}{4}\sech\left(\frac{mx}{2}\right)\nonumber\\
&&+\lambda A^3\left(- \frac{mx}{2}\tanh \left(\frac{mx}{2}\right)+1
\right)\sech^3\left(\frac{mx}{2}\right). \label{terzo} 
\eea
Inserting
\beq
f_2^{(3)}=cx\sech \left(\frac{mx}{2}\right)\tanh \left(\frac{mx}{2}\right)+c_2^{(3)}\sech\left(\frac{mx}{2}\right)
\eeq
into Eq.~(\ref{terzo}) and matching the inhomogeneous terms we find the two conditions
\beq
\frac{m^2A}{4}=cm\hsp \lambda A^3=2cm
\eeq
which are solved by
\beq
A=0,\ \pm \frac{m}{\sqrt{2\lambda}}.
\eeq
The first solution includes the unstable kink $f_2=0$ while the later two correspond to the stable kinks with
\bea
f_2^{(1)}&=&\pm\frac{m}{\sqrt{2\lambda}}\sech\left(\frac{mx}{2}\right)\\
f_1^{(2)}&=&-\frac{m^2x}{4\sqrt{2\lambda}} \ \sech^2\left(\frac{mx}{2}\right)\nonumber\\
f_2^{(3)}&=&\pm \frac{m^2 x}{4\sqrt{2\lambda}}\sech \left(\frac{mx}{2}\right)\tanh \left(\frac{mx}{2}\right)+c_2^{(3)}\sech\left(\frac{mx}{2}\right).\nonumber
\eea
Note that $c_2^{(3)}$, which shifts the normalization of $f_2$, is not yet constrained at this order, although in the exact solution $c_2^{(3)}=0$.  This is similar to the case of the normalization $A$ of $f_2^{(1)}$, which we just saw is fixed here at the third order.

\subsection{Order $O(\alpha^4)$}

At order $O(\alpha^4)$ we need only consider the first equation in (\ref{eom}).  The left hand side is $f_1^{(4)\prime\prime}(x)$.  Identifying it with the right hand side at $c_2^{(3)}=0$, we find
\bea
f_1^{(4)\prime\prime}(x)&=&m^2\left(1-\frac{3}{2}\sech^2\left(\frac{mx}{2}\right)\right)f_1^{(4)}(x)+\frac{3m^5x^2}{32\sqrt{2\lambda}}\sech^4\left(\frac{mx}{2}\right)\tanh\left(\frac{mx}{2}\right)\nonumber\\
&&+\frac{m^4x}{8\sqrt{2\lambda}}\left(2\sech^2\left(\frac{mx}{2}\right)-3\sech^4\left(\frac{mx}{2}\right)\right). \label{eq4}
\eea
We expand $f_1^{(4)}$ in powers of $x$
\beq
f_1^{(4)}=\frac{m^3}{\sqrt{2\lambda}}\left[ 
(mx)^2\alpha(x)+mx\beta(x)+\gamma(x)
\right]
\eeq
and match the terms with different powers of $x$ in (\ref{eq4}).  The $O(x^2)$ terms yield
\beq
\alpha\pp(x)=m^2\left(1-\frac{3}{2}\sech^2\left(\frac{mx}{2}\right)\right)\alpha(x)+\frac{3}{32}\sech^4\left(\frac{mx}{2}\right)\tanh\left(\frac{mx}{2}\right)
\eeq
which is solved by
\beq
\alpha(x)=-\frac{\sech^2\left(\frac{mx}{2}\right)\tanh\left(\frac{mx}{2}\right)}{16 m^2}. \label{alx}
\eeq
The order $O(x)$ equation
\beq
4m\alpha\p(x)+\beta\pp(x)=
m^2\left(1-\frac{3}{2}\sech^2\left(\frac{mx}{2}\right)\right)\beta(x)+\frac{\sech^2\left(\frac{mx}{2}\right)}{4}-\frac{3}{8}\sech^4\left(\frac{mx}{2}\right)
\eeq
turns out to be homogeneous once one substitutes in (\ref{alx}) and so one finds
\beq
\beta(x)=c_\beta \sech^2\left(\frac{mx}{2}\right)
\eeq
where $c_\beta$ is arbitrary.  Finally the order $O(x^0)$ equation
\beq
2m^2\alpha(x)+2m\beta\p(x)+\gamma\pp(x)=m^2\left(1-\frac{3}{2}\sech^2\left(\frac{mx}{2}\right)\right)\gamma(x)
\eeq
is homogeneous if
\beq
c_\beta=-\frac{1}{16m^2}
\eeq
so that one may set
\beq
\gamma(x)=0.
\eeq

Summarizing, we have found
\beq
f_1^{(4)}=-\frac{m^2}{16\sqrt{2\lambda}}x\sech^2\left(\frac{mx}{2}\right)\left[1+ 
mx\tanh\left(\frac{mx}{2}\right)
\right].
\eeq

Needless to say, this answer can be easily obtained by expanding the exact solution $f_2^\pm$ from Eq.~(\ref{es}) in $\alpha$.

\section{The Instanton} \label{instsez}

Let us repeat the above calculation but using the Ansatz
\beq
F_1(x,t)=F_1^{(0)}(x)+\alpha^2 F_1^{(2)}(x)A^2(\alpha t)\hsp
F_2(x,t)=\alpha F_2^{(1)}(x)A(\alpha t)+\alpha^3 F_2^{(3)}(x)c(\alpha t)
%F_2(x,t)=A(\alpha t) F_2(x)
\eeq
for some functions $A(\alpha t)$ and $c(\alpha t)$. %Here $f^{(n)}_{1,2} \equiv \alpha^n F^{(n)}_{1,2}$.
Now there are two dimensionless parameters, $\alpha$ which is small and $mt$ which ranges over all values.  This means, for example, that for some values of $t$, the combination $\alpha mt$ is large.  In fact, this region will have a fractional contribution to the instanton action of order unity, and so cannot be neglected.  Our strategy, therefore, will be to fix $t=\hat{t}$ and expand in $\alpha$.

With this caveat, the calculation proceeds identically to the case of the kink with $A$ replaced by $A(\alpha\hat{t})$ up to order $O(\alpha^2)$, yielding for example
\beq
F_1^{(2)}=-\frac{x\sl}{2\sqrt{2}}{\rm{sech}}^2
\left(\frac{mx}{2}\right).
\eeq
At order $O(\alpha^3)$ a contribution from the time derivatives first arises
\bea
\ddot A(\alpha\hat{t}) f^{(1)}_2(x)+c(\alpha\hat{t}) f^{(3) \pp}_2(x)&=&\frac{m^2 }{4}\left(1-2\sech^2 \left(\frac{mx}{2}\right)
\right)c(\alpha\hat{t}) f^{(3)}_2(x)-\frac{m^2 A(\alpha\hat{t})}{4}\sech\left(\frac{mx}{2}\right)\nonumber\\
&&+\lambda A^3(\alpha\hat{t})\left(- \frac{mx}{2}\tanh \left(\frac{mx}{2}\right)+1
\right)\sech^3\left(\frac{mx}{2}\right)\nonumber\\
&=&\ddot A(\alpha\hat{t}) \sech \left(\frac{mx}{2}\right) +c(\alpha\hat{t}) f^{(3) \pp}_2(x).
\eea
 Here the dot is a derivative with respect to the argument $\alpha t$, evaluated at $t=\hat{t}$.   
 
Subtracting the new term from both sides
\bea
c(\alpha\hat{t}) f^{(3) \pp}_2(x)&=&\frac{m^2 c(\alpha\hat{t})}{4}\left(1-2\sech^2 \left(\frac{mx}{2}\right)
\right)f^{(3)}_2(x)-\left[\frac{m^2 A(\alpha\hat{t}) }{4}+\ddot A(\alpha\hat{t}) \right]\sech\left(\frac{mx}{2}\right)\nonumber\\
&&+\lambda A^3(\alpha\hat{t})\left( -\frac{mx}{2}\tanh \left(\frac{mx}{2}\right)+1
\right)\sech^3\left(\frac{mx}{2}\right).
\eea
Choosing
\beq
F_2^{(3)}(x)=x\sech \left(\frac{mx}{2}\right)\tanh \left(\frac{mx}{2}\right)
\eeq
and matching the inhomogeneous terms again, we now find
\beq
\frac{m^2A(\alpha\hat{t})}{4}+\ddot A(\alpha\hat{t})=c(\alpha\hat{t})m\hsp \lambda A^3(\alpha\hat{t})=2c(\alpha\hat{t})m.
\eeq
Therefore $c(\alpha\hat{t})m=\lambda A^3(\alpha\hat{t})/2$ and so
\beq
\ddot A(\alpha\hat{t})=\frac{\lambda A^3(\alpha\hat{t})}{2}-\frac{m^2 A(\alpha\hat{t})}{4}.
\eeq
Demanding that this equation be satisfied at all $t=\hat{t}$ and imposing the boundary condition (\ref{tbc}) we find
\beq
A(\alpha t)=\frac{m}{\sqrt{2\lambda}}\tanh\left(\alpha\frac{m t}{2\sqrt 2}\right).
\eeq
We conclude that the instanton solution is
\bea
F_1(x,t)&=&\frac{m}{\sqrt{2\lambda}}{\rm{tanh}}\left(\frac{mx}{2}\right)-\alpha^2\frac{m^2x}{4\sqrt{2\lambda}}\sech^2\left(\frac{mx}{2}\right)\tanh^2\left(\alpha\frac{m t}{2\sqrt 2}\right)+O(\alpha^4)\nonumber\\
F_2(x,t)&=&\alpha\frac{m}{\sqrt{2\lambda}}\sech\left(\frac{mx}{2}\right)\tanh\left(\alpha\frac{m t}{2\sqrt 2}\right)\nonumber\\
&&
+\alpha^3\frac{m^2x}{4\sqrt{2\lambda}}\sech\left(\frac{mx}{2}\right)\tanh\left(\frac{mx}{2}\right)\tanh^3\left(\alpha\frac{m t}{2\sqrt 2}\right)
+O(\alpha^5).
\eea
This may be obtained from the kink solution with the replacement
\beq
\alpha \rightarrow \alpha\tanh\left(\alpha\frac{m t}{2\sqrt 2}\right). \label{rimp}
\eeq
As a result, everything evolves in phase, allowing for the cancellation of the instanton Lagrangian with the kink Lagrangian, and so of the Euclidean action at this order.  We conclude that, at this order, there is not yet any splitting between the Hamiltonian eigenstates in the kink sector.  We will need to press on to $O(\alpha^4)$.
%\red{This subsection was not done carefully at all ... there could very well be critical mistakes.  For example, I haven't justified going to all orders in the $\alpha$ expansion of the time dependence but not the space dependence.}

Let us write the instanton solution at $O(\alpha^4)$ as
\bea
F_1^{(4)}&=&f_1^{(4)}\tanh^4\left(\alpha\frac{m t}{2\sqrt 2}\right)+\tilde F_1^{(4)} \label{in4}\\
f_1^{(4)}&=&-\frac{m^2}{16\sqrt{2\lambda}}x\sech^2\left(\frac{mx}{2}\right)\left[1+ 
mx\tanh\left(\frac{mx}{2}\right)
\right]\nonumber
\eea
where the $f_1^{(4)}$ term is obtained from the kink solution via (\ref{rimp}) and the other term is the correction, to which we now turn our attention.  

We then substitute this into the first Euclidean equation of motion at order $O(\alpha^4)$.  The first term on the right hand side of (\ref{in4}) satisfied this equation in the time independent case, and so the time derivative term on the left must cancel the $\tilde F_2^{(4)}$ contributions
\bea
\tilde F_1^{(4)\prime\prime}+\frac{\ddot F^{(2)}_1\big|_{O(\alpha^2)}}{\alpha^2}=\lambda \tilde F_1^{(4)}\left(3F_1^{(0)2}-\frac{m^2}{2\lambda}\right)&=&m^2\left(1-\frac{3}{2}\sech^2\left(\frac{mx}{2}\right)\right)\tilde F_1^{(4)}. \label{f14eq}
\eea
Using
\beq
F_1^{(2)}=-\frac{m^2x}{4\sqrt{2\lambda}}\sech^2\left(\frac{mx}{2}\right)\tanh^2\left(\alpha\frac{m t}{2\sqrt 2}\right)
\eeq
we find
\beq
\frac{\ddot F^{(2)}_1\big|_{O(\alpha^2)}}{\alpha^2}=\frac{m^4x}{16\sqrt{2\lambda}}\sech^2\left(\frac{mx}{2}\right)\left[2\sech^2\left(\alpha\frac{m t}{2\sqrt 2}\right) -3\sech^4\left(\alpha\frac{m t}{2\sqrt 2}\right)
\right].
\eeq

Inserting this into Eq.~(\ref{f14eq}) we obtain an equation of the form
\beq
L\pp(x)=\left( 4-6\sech^2(x)\right)L(x)+x\sech^2(x).
\eeq
The solution to this equation is
\bea
L(x)&=&\frac{\tanh(x)}{24}+\frac{x}{24}\left[ 
4\cosh^2(x)+4-5\sech^2(x)\right]
+\frac{\sech^2(x)}{16}\left[{\mathrm{Li}}_2(-e^{2x})- {\mathrm{Li}}_2(-e^{-2x})
\right]\nonumber\\
&&
-\frac{{\mathrm{ln}}\left(e^x+e^{-x}\right)}{12}\left[2\sinh(x)\cosh(x)+3\tanh(x)\right].
\eea
We thus conclude
\beq
\tilde F_1^{(4)}=\frac{m}{2\sqrt{2\lambda}}L\left(\frac{mx}{2}\right)\left[3\sech^4\left(\alpha\frac{m t}{2\sqrt 2}\right)-2\sech^2\left(\alpha\frac{m t}{2\sqrt 2}\right)
\right].
\eeq

%\red{}
%\beq
%F_1^{(4)}(x,t)=-\frac{m^2}{16\sqrt{2\lambda}}x\sech^2\left(\frac{mx}{2}\right)\left[1+ 
%mx\tanh\left(\frac{mx}{2}\right)
%\right]\tanh^4\left(\alpha\frac{m t}{2\sqrt 2}\right)+\tilde{F}_1^{(4)}(x,t).
%\eeq
%\beq
%\frac{\ddot F_1^{(2)}}{\alpha^2}=-\tilde{F}_1^{(4)\prime\prime}+
%m^2\left(1-\frac{3}{2}\sech^2\left(\frac{mx}{2}\right)\right)\tilde{F}_1^{(4)}
%\eeq

\begin{figure}[htbp]
\centering
\includegraphics[width = 0.8\textwidth]{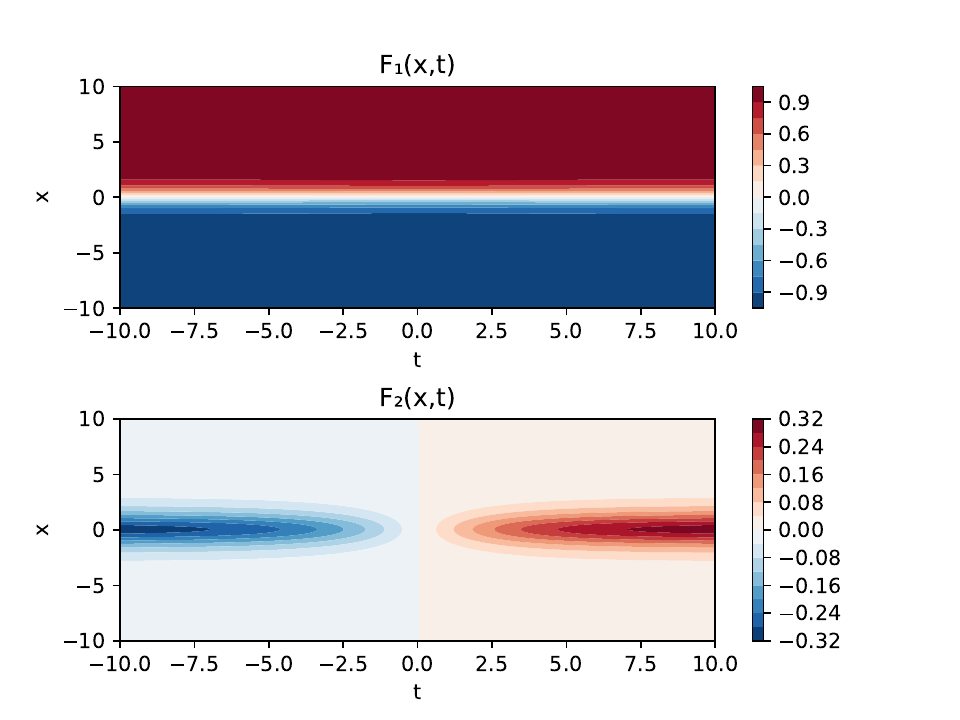}
\includegraphics[width = 0.8\textwidth]{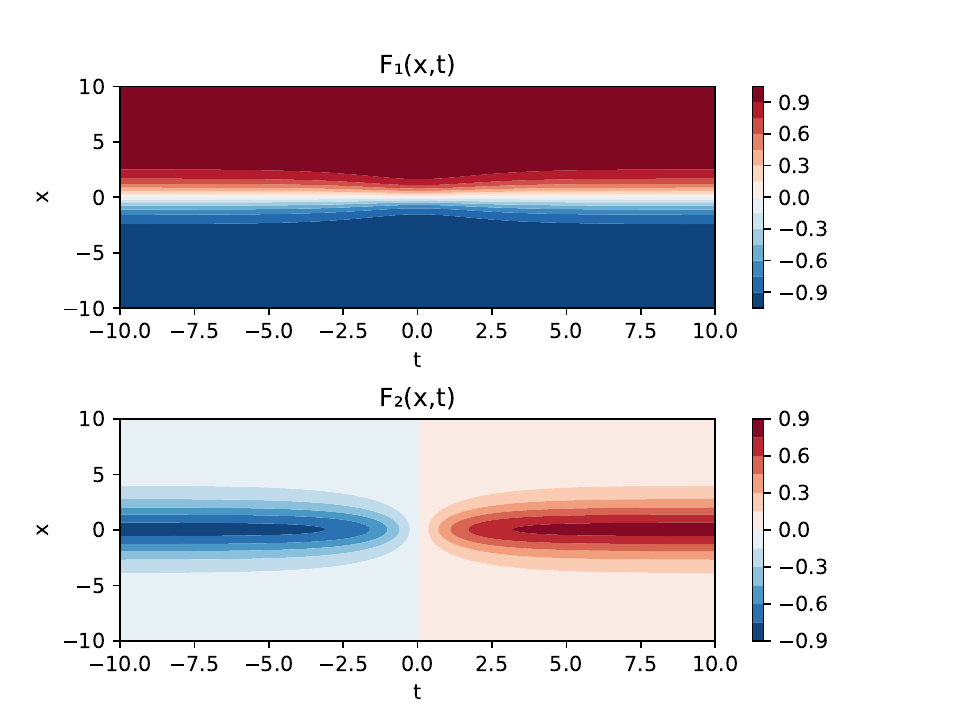}
\caption{The field configurations for the instanton at $\alpha=0.3$ (top) and $\alpha=0.8$ (bottom), which have Euclidean actions $\Delta S_E=0.052$ and $1.42$.  Here $m=\lambda=2$.}\label{alfig}
\end{figure}

Numerically, we have checked that the instanton exists for higher values of $\alpha$, and we have constructed the solutions using a gradient descent method. Indeed, the fact that the solutions $f_i^\pm(x)$ are degenerate in energy implies that the two-dimensional instanton that connects them does not present negative modes in its perturbation spectrum, hence it should be a global minimum of the Euclidean action (as opposed to other instanton (bounce) solutions in models with non-degenerate minima, see e.g. \cite{Chigusa:2019wxb}).  Therefore, the instanton can be constructed using a standard gradient flow algorithm. We have implemented such an algorithm on a discrete grid of $N\times N$ points, with $N=400$, which allowed us to compute the corresponding Euclidean action with less than one percent error.

In Fig.~\ref{alfig} we plot the configurations at $\alpha=0.3$ where our perturbative approach works quite well, and also $\alpha=0.8$ which is beyond the range of our $\alpha$ expansion.  

\section{The Energy Splitting} \label{azsez}

\subsection{The Expansion}

The Euclidean action $S_E$ is the time integral of the Euclidean Lagrangian, which is equal to the Hamiltonian.  This in turn is a space integral of the Euclidean Lagrange density, which is equal to the Hamiltonian density.  We will decompose these as
\beq
S_E=\int dt\int dx \mathcal{L}_E\hsp \mathcal{L}_E=T+U
\eeq
where $T$ and $U$ are the kinetic and potential terms in the Hamiltonian density
\beq
T=\frac{\sum_{a=1}^2\left(\partial_t{F}_a\partial_t{F}_a+\partial_x F_a\partial_x F_a\right)}{2}\hsp
U=\frac{\lambda\left(F_1^2+F_2^2-\frac{m^2}{2\lambda}\right)^2}{4}+\frac{(1-\alpha^2)m^2}{8}F_2^2
\eeq
which we will expand in powers of $\alpha$
\beq
T=\sum_n \alpha^n T_n\hsp U=\sum_n \alpha^n U_n.
\eeq

The Euclidean action enjoys a long distance divergence, resulting from the fact that at $|t|\rightarrow\infty$ the instanton tends to a kink which has a nonzero Lagrangian.  However, the quantity that appears in the instanton gas approximation to the energy splitting is not $S_E$, which depends on the physically irrelevant zero-point of the Hamiltonian density, but rather the difference between the Euclidean action of the instanton and that of the kink.  This difference in infinities is well defined because one subtracts the Lagrangians before performing the $t$ integration. 

In other words, the mass splitting will depend on the quantity
\beq
\Delta S_E=\int dt\int dx \Delta\mathcal{L}_E\hsp \Delta\mathcal{L}_E=T+U-T^K-U^K
\label{deltS}
\eeq
where $T^K$ and $U^K$ are the kinetic and potential energy densities of the kink.  
We will also decompose these in powers of $\alpha$
\beq
T^K=\sum_n \alpha^n T^K_n\hsp U^K=\sum_n \alpha^n U^K_n. \label{eseq}
\eeq
An alternative but equivalent way to understand this  is to consider the instanton in the MSTB model as a static configuration formed by a vortex attached to a domain wall string in a 2+1 dimensional version of the theory. Then, the associated Euclidean action can be thought of as the total energy of such configuration, which diverges due to the infinite length of the wall. However, the presence of the vortex alters the total energy of the configuration, and its energy relative to the case of a domain wall without vortex is precisely \eqref{deltS}.

\subsection{The Instanton's Action}

\subsection{Order $O(\alpha^0)$}

At leading order these quantities only depend on $F_1^{(0)}$, which, as stated in Eq.~(\ref{ordz}), is the usual kink from the $\phi^4$ theory.  As a result, these quantities are those of the $\phi^4$ theory
\beq
T_0=U_0=T^K_0=T^U_0=\frac{m^4}{16\lambda}\sech^4\left(\frac{mx}{2}\right).
\eeq

We note that $T=U$, as the kink and so the instanton is BPS at this order.  It is inversely proportional to $\lambda$, and so would lead to a nonperturbative $e^{-1/\lambda}$ type splitting.  However, as 
\beq
T_0+U_0-T^K_0-U^K_0=0
\eeq
the contribution to the energy splitting at this order vanishes.  Of course this is of no surprise, because at $\alpha=0$ there is only one kink solution and so there is no splitting.

\subsection{Order $O(\alpha^2)$}

At the next order we find
\bea
T_2&=&\partial_x F_1^{(0)}\partial_x F_1^{(2)}+\frac{1}{2}\partial_x F_2^{(1)}\partial_x F_2^{(1)}\\
&=&\frac{m^4}{16\lambda}\left[ 
\left(mx\tanh\left(\frac{mx}{2}\right)-2\right)\sech^4\left(\frac{mx}{2}\right)+\sech^2\left(\frac{mx}{2}\right)\right]\tanh^2\left(\alpha\frac{mt}{2\sqrt{2}}
\right)\nonumber\\
U_2&=&\frac{\lambda}{2}\left(F_1^{(0)2}-\frac{m^2}{2\lambda}\right)\left(2F_1^{(0)}F_1^{(2)}+F_2^{(1)2}\right)+\frac{m^2}{8}F_2^{(1)^2}=T_2.\nonumber
\eea
Again, $T_2=U_2$ as the instanton is apparently BPS at this order as well \cite{adambps}.  Integrating over $x$, the quantities $T_2$ and $U_2$ both vanish at each value of $t$.  

The corresponding calculation for the kink is identical, one need only take the limit $|t|\rightarrow\infty$ so that the time-dependent $\tanh$ factors become unity.  Again, the $x$ integration leads to zero.  So both Euclidean Lagrangians vanish and do not contribute to $\Delta S_E$ at this order.

\subsection{Order $O(\alpha^4)$}

At the next order the instanton's kinetic and potential Lagrangian densities are
\bea
T_4&=&\frac{\partial_t F_2^{(1)}\partial_t F_2^{(1)}}{2\alpha^2}+\partial_x F_1^{(0)}\partial_x F_1^{(4)}+\partial_x F_2^{(1)}\partial_x F_2^{(3)}+\frac{\partial_x F_1^{(2)}\partial_x F_1^{(2)}}{2}\\
U_4&=&\frac{\lambda}{2}\left(F_1^{(0)\ 2}-\frac{m^2}{2\lambda}\right)\left(2F_1^{(0)}F_1^{(4)}+2F_2^{(1)}F_2^{(3)}+F_1^{(2)\ 2}\right)+\frac{\lambda}{4}\left(2F_1^{(0)}F_1^{(2)}+F_2^{(1)\ 2}\right)^2\nonumber\\
&&+\frac{m^2}{4}F_2^{(1)}F_2^{(3)}-\frac{m^2}{8}F_2^{(1)\ 2}.\nonumber
\eea
In the case of the kink there are no time derivatives and so one finds
\bea
T_4^K&=&\partial_x f_1^{(0)}\partial_x f_1^{(4)}+\partial_x f_2^{(1)}\partial_x f_2^{(3)}+\frac{\partial_x f_1^{(2)}\partial_x f_1^{(2)}}{2}\\
U_4^K&=&\frac{\lambda}{2}\left(f_1^{(0)\ 2}-\frac{m^2}{2\lambda}\right)\left(2f_1^{(0)}f_1^{(4)}+2f_2^{(1)}f_2^{(3)}+f_1^{(2)\ 2}\right)+\frac{\lambda}{4}\left(2f_1^{(0)}f_1^{(2)}+f_2^{(1)\ 2}\right)^2\nonumber\\
&&+\frac{m^2}{4}f_2^{(1)}f_2^{(3)}-\frac{m^2}{8}f_2^{(1)\ 2}.\nonumber
\eea

Using the relations
\bea
F_1^{(0)}=f_1^{(0)}\hsp 
F_2^{(1)}=f_2^{(1)}\tanh\left(\alpha\frac{mt}{2\sqrt{2}}
\right)\hsp 
F_1^{(2)}=f_1^{(2)}\tanh^2\left(\alpha\frac{mt}{2\sqrt{2}}
\right)\\
F_2^{(3)}=f_2^{(3)}\tanh^3\left(\alpha\frac{mt}{2\sqrt{2}}
\right)\hsp
F_1^{(4)}=f_1^{(4)}\tanh^4\left(\alpha\frac{mt}{2\sqrt{2}}
\right)+\tilde{F}_1^{(4)}\nonumber
\eea
we find the relations
\bea
T_4&=&T_4^K\tanh^4\left(\alpha\frac{mt}{2\sqrt{2}}
\right)+\frac{\partial_t F_2^{(1)}\partial_t F_2^{(1)}}{2\alpha^2}+\partial_x F_1^{(0)}\partial_x \tilde{F}_1^{(4)}\label{deceq}\\
U_4&=&U_4^K\tanh^4\left(\alpha\frac{mt}{2\sqrt{2}}
\right)+\lambda \left(F_1^{(0)\ 2}-\frac{m^2}{2\lambda}\right)F_1^{(0)}\tilde{F}_1^{(4)}\nonumber\\
&&+\frac{m^2}{8}f_2^{(1)\ 2}\left(
\tanh^4\left(\alpha\frac{mt}{2\sqrt{2}}
\right)
-\tanh^2\left(\alpha\frac{mt}{2\sqrt{2}}
\right)
\right).
\nonumber
\eea

Inserting the BPS kink solution one finds
\bea
T^K_4&=&U^K_4=\frac{m^4}{128\lambda}\left[ 
8\sech^2\left(\frac{mx}{2}\right)-8+mx\tanh\left(\frac{mx}{2}\right)\left(4-14\sech^2\left(\frac{mx}{2}\right)\right)\right.\nonumber\\&&\left.+m^2x^2\left(4\sech^2\left(\frac{mx}{2}\right)-5\sech^4\left(\frac{mx}{2}\right)\right)
\right]\sech^2\left(\frac{mx}{2}\right).
\eea

The terms proportional to $T^K$
\beq
(T_4^K+U_4^K)\tanh^4\left(\alpha\frac{mt}{2\sqrt{2}}
\right)-T_4^K-U_4^K=2 T_4^K\left[\sech^4\left(\alpha\frac{mt}{2\sqrt{2}}
\right)-2\sech^2\left(\alpha\frac{mt}{2\sqrt{2}}
\right)\right]
\eeq
yield a finite contribution to the Euclidean action
\bea
\int dx\int dt (T_4^K+U_4^K)\tanh^4\left(\alpha\frac{mt}{2\sqrt{2}}
\right)-T_4^K-U_4^K&=&2\left(\int dx T_4^k\right)\label{kcon}\\
&&\hspace{-8cm}\ \ \ \times \int dt\left[ \sech^4\left(\alpha\frac{mt}{2\sqrt{2}}
\right)-2\sech^2\left(\alpha\frac{mt}{2\sqrt{2}}
\right)\right]\nonumber\\
&&\hspace{-8cm}=\frac{m^4}{64\lambda}\left(-\frac{8}{m}
\right)
\left(-\frac{16\sqrt{2}}{3\alpha m}
\right)=\frac{4m^2}{3\sqrt{2}\lambda\alpha}.\nonumber
\eea
We recall that $\lambda/m^2$ is our dimensionless coupling parameter and that the contribution to $\Delta S_E$ is this multiplied by $\alpha^4$, as we have factored this out in (\ref{eseq}).

\begin{figure}[htb!]
    \centering
    \includegraphics[width=0.7\linewidth]{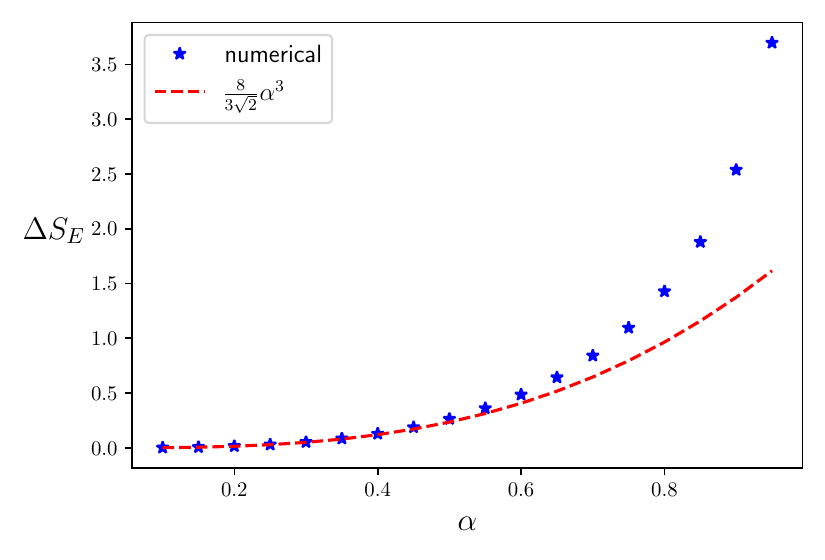}
    \caption{Relative Euclidean action as a function of the coupling $\alpha$, in the case $m=\lambda=2$.}
    \label{fig:S_alpha}
\end{figure}

We have now evaluated the contribution to the action from the first term in both expressions of Eq.~(\ref{deceq}).  We will now evaluate the contributions arising from the other two terms in each equation.  The first contribution from $T_4$ is
\beq
\int dx\int dt \frac{\partial_t F_2^{(1)}\partial_t F_2^{(1)}}{2\alpha^2}= \frac{m^4}{32\lambda}\int dx \sech^2\left(\frac{mx}{2}\right)\int dt \sech^4\left(\alpha\frac{mt}{2\sqrt{2}}\right)=\frac{\sqrt{2}m^2}{3\lambda\alpha} \label{t1}
\eeq
while the second is
\bea
\int dx\int dt \partial_x F_1^{(0)}\partial_x \tilde{F}_1^{(4)}&=&\frac{m^4}{16\lambda}\int dx\ \sech^2 \left(\frac{mx}{2}\right)L^\prime\left(\frac{mx}{2}\right)\\
&&\times
\int dt \left[3\sech^4\left(\alpha\frac{m t}{2\sqrt 2}\right)-2\sech^2\left(\alpha\frac{m t}{2\sqrt 2}\right)
\right]=0.\nonumber
\eea
This last contribution vanished because the $t$ integration gave zero.  The first contribution from $U_4$ is
\bea
\int dx\int dt \lambda \left(F_1^{(0)\ 2}-\frac{m^2}{2\lambda}\right)F_1^{(0)}\tilde{F}_1^{(4)}&=&-\frac{m^4}{8\lambda}\int dx\sech^2\left(\frac{mx}{2}\right)\tanh\left(\frac{mx}{2}\right)L\left(\frac{mx}{2}\right)\nonumber\\
&&\hspace{-2cm}\times\int dt \left[3\sech^4\left(\alpha\frac{m t}{2\sqrt 2}\right)-2\sech^2\left(\alpha\frac{m t}{2\sqrt 2}\right)
\right]=0
\eea
which also vanishes because the time dependence in $\tilde{F}_1^{(4)}$ integrates to zero.  The last contribution is
\bea
\int dx\int dt \frac{m^2}{8}f_2^{(1)\ 2}\left(
\tanh^4\left(\alpha\frac{mt}{2\sqrt{2}}
\right)
-\tanh^2\left(\alpha\frac{mt}{2\sqrt{2}}
\right)
\right)&=&\frac{m^4}{16\lambda}\int dx\sech^2\left(\frac{mx}{2}\right)\nonumber\\
&&\hspace{-10
cm}\times\int dt\left(
\sech^4\left(\alpha\frac{mt}{2\sqrt{2}}
\right)
-\sech^2\left(\alpha\frac{mt}{2\sqrt{2}}
\right)\right)=-\frac{\sqrt{2}m^2}{3\lambda\alpha}
\eea
which cancels the contribution (\ref{t1}) leaving only (\ref{kcon}).  We therefore conclude
\beq
\Delta S_E=\frac{4m^2\alpha^3}{3\sqrt{2}\lambda}+O(\alpha^5).
\eeq
Note that at $\alpha=0.3$ this is in good agreement with the $\Delta S_E$ that we have found numerically by directly solving the Euclidean equations of motion \eqref{eom}, as reported in Fig.~\ref{fig:S_alpha}.

\subsection{The Instanton Gas Approximation}

When $\alpha>0$, there is a single topological sector corresponding to kinks in the MSTB models.  It includes kinks with both $f_2\geq 0$ and also $f_2\leq 0$, or more precisely the solutions $f^+$ and $f^-$.  Furthermore, these kinks may be centered at an arbitrary point $x_0$.  Altogether this moduli space consists of two real lines.

The key observation in this note is that the instanton that interpolates between the kink $f^+$ and the kink $f^-$ enjoys a finite Euclidean action, or more precisely a finite action difference $\Delta S_E$, in the case in which both are centered at the same value of $x_0$.  One may be concerned about instanton transitions between $f^+$ and $f^-$ kinks at different values of $x_0$.  These necessarily have higher actions $\Delta S_E$ and we intend to return to them in the future.  For now we will simply ignore them, hoping that their contribution is subleading.

With this caveat, our moduli space of kinks is reduced to the double-well model of Refs.~\cite{polyakovdw,abc,hannay24}, albeit with a more complicated excitation spectrum.  Following the argument there, one expects that the true ground state will consist of wave functions that are symmetric with respect to the $f^+\leftrightarrow f^-$ symmetry and a first excited state which is odd.  The energy splitting at leading order\footnote{Here the leading order contribution is the term linear in the action that appears in the exponential, whereas subleading orders include multiplicative powers of the action which are not exponentiated.} will be proportional to
\beq
\Delta E\propto e^{-\Delta S_E}.
\eeq
While the excitation spectrum is more complicated in the present case, this only contributes to the splitting at the subleading order, and we will turn to this problem in future work.  

This treatment also suggests that the zero mode relating distinct values of $x_0$ may be treated by considering a two-dimensional quantum mechanical double-well problem, in which the potential is independent of the second dimension.  In this case, it is clear that the ground state and first excited wavefunctions of the two-dimensional problem are just those of the one-dimensional problem, and they are independent of the second dimension.  As the derivatives of the wave functions in the second dimension vanish, the energies are identical to those of the one-dimensional problem and so the energy splitting is as in the one-dimensional problem, despite the presence of an additional zero mode represented by the translation in the second dimension.  This is not to suggest that the energy splitting will be the same at all orders, the two-dimensional quantum mechanical model only captures the additional zero mode and not the more complicated normal modes, which contribute to the energy splitting at the next order.  

In summary, we feel that we have strong arguments to suggest that this additional zero mode does not affect the leading order energy splitting, and indeed in the literature we know of no case in which a zero mode affects the leading order mass splitting.  However in the future we intend to turn to the next order, which depends on all perturbations including the zero modes for the position $x_0$ of the kink and the time $t_0$ of the instanton.

%As the tunneling probability falls off exponentially with the angle of the deviation if the tunneling from the perpendicular between the wells, we suspect that the tunneling will be dominates by paths close to the perpendicular, and so the exponential appearing in the splitting will be $-S_E$ just as in the case of the one-dimensional double-well.  Indeed, in the instanton literature there are many cases of solitons with collective coordinates, and to our knowledge in no case does a zero mode lead to a finite shift in an energy splitting at leading order.

\section{What Next?}

The heuristic argument in the previous paragraph becomes precise in the Hamiltonian formulation of the problem.  Motivated by Bender and Wu's application of WKB techniques to calculate higher order corrections to the anharmonic oscillator in Refs.~\cite{bw1,bw2}, Banks, Bender and Wu introduced a higher-dimensional WKB approach in Refs.~\cite{bbw,bb}.  The key to this approach was the reduction of the intractible multidimensional problem to a one-dimensional problem through the realization that the wave function is localized near discrete most probable trajectories.  This approach was in turn generalized to quantum field theories in Ref.~\cite{gswkb} which explained that these trajectories are the solutions of the Euclidean equations of motion, and thus the results are equivalent to those of an instanton gas.  The two-dimensional quantum mechanics of the previous paragraph is simply a projection of the Hilbert space onto the kink moduli space, which consists of two lines, plus the plane of least Euclidean action that connects them.  Bender, Wu and Banks have shown that the wave function is supported on this surface with exponential fall-off in other directions, and that the leading order WKB approximation depends only on the Euclidean action on this surface, $S_E$.  

While that work was largely formal and crouched in the language of quantum mechanics, an application to quantum field theory without \cite{vega1} and with \cite{vega2} zero modes soon followed together with de Vega, with key steps described such as the matching conditions between the basins of the two minima. 

The above approach directly describes the quantum states, and so has more applications than the instanton gas approach to mass corrections.  Thus in the future we would like to use the instanton gas results presented here to guide an application of de Vega's approach to the MSTB model.  

The final formula for the kink mass will be obtained by summing the entire trans-series.  Such a program may be possible thanks to recent developments described in Refs.~\cite{trans24,shu24}.

Needless to say, we also will use it as a guide to instanton corrections that arise in a sector corresponding to a single solution, be it a vacuum or a soliton ground state, that arise from brief excursions of the fields into another basin.  These will be present in all models with multiple vacua, not just those lucky models with multiple solitons in the same topological sector.

\section* {Acknowledgement}

\noindent
We are grateful to Alberto Alonso Izquierdo for insights into the MSTB model that motivated this work.  AGMC wants to thank J.J. Blanco-Pillado for insightful discussions.
JE is supported by NSFC MianShang grants 11875296 and 11675223 and by the Ministry of Education, Science, Culture and Sport of the Republic of Armenia under the Remote Laboratory Program, grant number 24RL-1C047.  AW is supported by the Spanish MCIN with funding from European Union
NextGenerationEU (PRTRC17.I1) and Consejeria de Educacion from JCyL through the
QCAYLE project, as well as the MCIN project PID2020-113406GB-I0. AGMC acknowledges support from the PID2021-123703NB-C21
grant funded by MCIN/AEI/10.13039/501100011033/and by ERDF;“ A way of making Europe”;
and the Basque Government grant (IT-1628-22).

\end{document}